# STUDY OF GAMMA CASCADES OF $^{59}$Ni BY THERMAL NEUTRON REACTION

Nguyen An Son[1], Pham Dinh Khang[2], Nguyen Duc Hoa[1], Nguyen Xuan Hai[3], Nguyen Thi Minh Sang[1]

[1] University of Dalat, 01 Phu Dong Thien Vuong, Dalat, Vietnam
[2] Nuclear Training Center, 140 Nguyen Tuan, Hanoi, Vietnam
[3] Dalat nuclear Research Institute, 01 Nguyen Tu Luc, Dalat, Vietnam

**Abstract:** The quantum properties are important to study nuclear structure. The energy, spin, parity, transition order are usually interesting to research. In this experiment, $^{59}$Ni is activated by thermal neutron on 3$^{rd}$ horizontal channel of Dalat nuclear Reactor, to get data by "event-event" gamma-gamma coincidence. This report shown two-step cascades energies, relative intensities, spin, parity and electromagnetic transition probabilities of $^{59}$Ni by $^{58}$Ni(n$^{th}$, 2γ)$^{59}$Ni reaction.

**Key words:** $^{A}$X(n$_{th}$, 2γ)$^{A+1}$X reaction; quantum properties.

## 1. INTRODUCTION

The study of gamma decay of $^{59}$Ni have been previously published in many works based on different reactions such as (α, n) with energy smaller than 2 MeV [3], the (p, γ) reaction with energy range from 3 to 6 MeV by the work of M. Pichevar and colleagues [5]. The energy and the intensity of $^{59}$Ni gamma transfers from B$_n$ to ground state as showed in the report of IAEA [2].

However, the results of previous studies didn't show the agreement of the experimental data with and nuclear theory models, and there are many unclear problems in theoretical and empirical nuclear structure.

In this experiment, we studied $^{59}$Ni cascade gamma transfer base on gamma - gamma coincidence system with "event-event" method on 3$^{rd}$ horizontal channel of Dalat nuclear Reactor. From the experiment data, we applied single particle nuclear model to determine some quantum properties and electromagnetic transfer probabilities.

## 2 . EXPERIMENT AND METHOD

### 2.1 Samples and experimental systems

Experimental sample is natural nickel. The isotope ratio of the nickel samples and thermal neutron capture cross section are $^{58}$Ni (68.0769 % and 4.6 barn); $^{60}$Ni (26,2231%, 2,9 barn); $^{61}$Ni (2,5%, 1,1399 barn); $^{62}$Ni (14,5%, 3,6345 barn); $^{64}$Ni (0,9256%, 1,58 barn) respectively [1].

The neutron beam, sample and detector was set up for maximum efficiency of gamma detection. In this experiment the sample is set at 45$^o$ from neutron beam, two detectors are placed opposite (180$^o$) with each other. The thermal neutron flux at sample position was about 10$^6$ n/cm$^2$/s. Cadmium coefficient is 900 (1 mm in thickness).

Experimental system was the gamma – gamma coincidence with event-event counting method, as showed in Fig. 1. The operating principle was briefly described as follow: The signal from two detectors was amplified and shaped by the amplifier (Amp. 7072A), which convert the output signal from the amplifier to digital signal when the conditions of 7811R interface is satisfied. Timing signals from two detectors was amplified and shaped by fast amplifier (FFA 474). The output signals from FFA 474 went through the Constant Fraction

Discriminator (CFD 584) and then the output signals of two CFD came to Start and Stop gates of TAC with one output signal was delayed correlatively with the time of coincidence gates. The valid output of TAC was the control signal for ADC 7027 which decide whether or not accept the signal conversion. The output of TAC is digital signal which was converted from the analog signal by ADC 8713 contain the timing information.

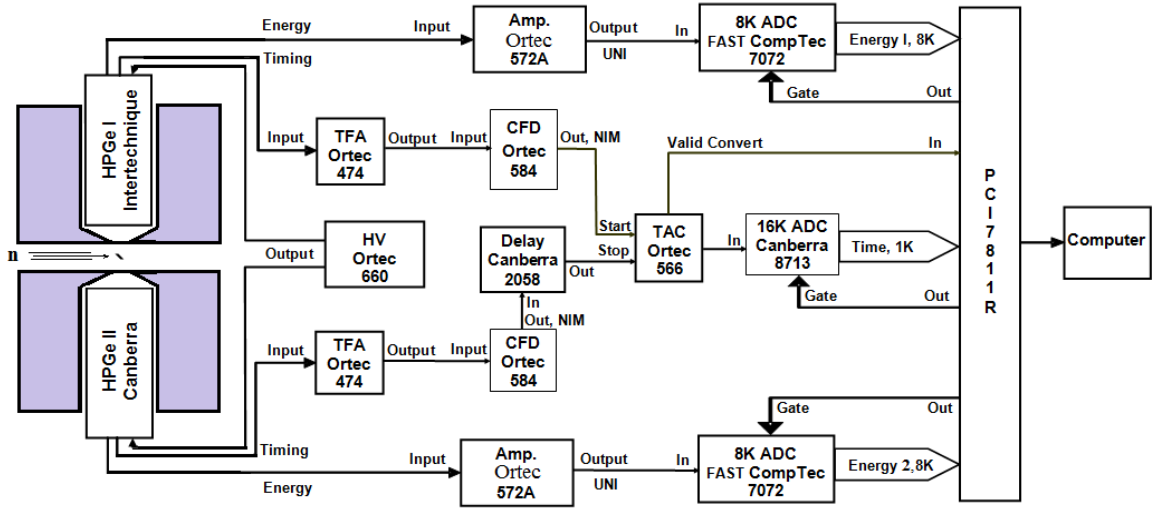

Fig. 1. Experimental system [6].

## 2.2 Quantum properties

Assuming that $J^\pi$ spin and parity are the ground state of nuclear, the spin and the parity of the compound nuclear when capturing neutron (s wave neutron capture) are ability $J^\pi \pm \tfrac{1}{2}$. Because of the lifetime of nuclei at excited states is very short, gamma radiations emitted from compound nuclear are usually electric dipole radiations E1, magnetic dipole M1, electric quadruple E2 or a mixture of M1 + E2. Selection rules for the multiple order of radiation are identified by:

$$|J_i - J_f| \leq L \leq J_i + J_f \tag{1}$$

L is multiple order. $J_i$ is the spin of the initial state, $J_f$ is the spin of the final state.

In this experiment, the relative intensity of cascade gamma transfer is calculated:

$$I_i^{\gamma-\gamma} = \frac{S_i^{\gamma-\gamma}}{\sum_{1}^{n} S_i^{\gamma-\gamma}} \tag{2}$$

$S_i^{\gamma-\gamma}$ is the calibrated area of i[th] peak in the cascade transfer.

According to the shell model, the transferred probability is [4]:

Electric transfer probability: $$T_\gamma^{EL} = \frac{8\pi(L+1)e^2 b^L}{L[(2L+1)!!]^2 \hbar} \left(\frac{E_\gamma}{\hbar c}\right)^{2L+1} B(EL) \tag{3}$$

Magnetic transfer probability $$T_\gamma^{ML} = \frac{8\pi(L+1)\mu_N^2 b^{L-1}}{L[(2L+1)!!]^2 \hbar} \left(\frac{E_\gamma}{\hbar c}\right)^{2L+1} B(ML) \tag{4}$$

Where ℏ is the Dirac constant, L is the multiple order of the gamma radiation, E is the energy of gamma radiation (keV), R is the nuclear radius, $e^2$= 1.440×10$^{-10}$ keVcm, $\mu_N^2$ = 1.5922×10$^{-38}$ keVcm$^3$, and b = 10$^{-24}$cm$^2$.

## 3. RESULT AND DISCUSSION
### 3.1. Energy, relative intensity, spin, the intermediate level of two-step cascade transfer

The time for Ni sample measurement was about 400 hours. Fig 2 is a part of sum spectrum of $^{59}$Ni.

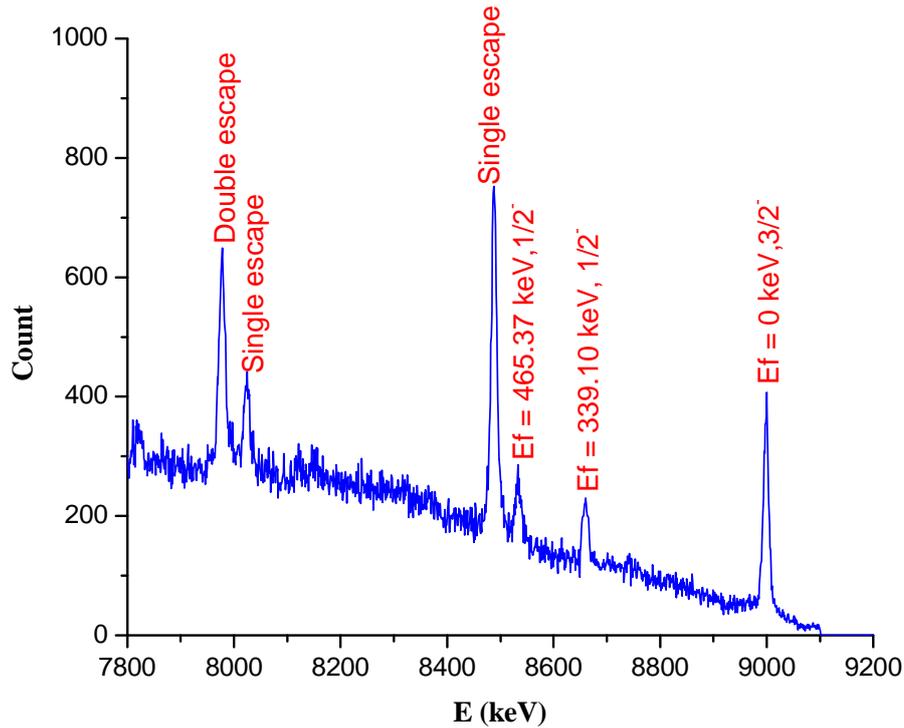

Fig. 2. A part of sum spectrum.

$^{58}$Ni is even – even nuclear including 28 protons and 30 neutrons. At ground state, the spin and parity of $^{58}$Ni is 0$^+$ and $^{59}$Ni is 3/2$^-$. When $^{58}$Ni captured neutron and became a compound nuclear that have spin and parity of 1/2$^+$ [2][5]. Table 1 presents some experimental results.

Table 1. Some experimental data obtained from the $^{58}$Ni (n , 2γ) $^{59}$Ni reaction.

| E1(keV) | E2(keV) | E$_L$(keV) | Experimental Spin | Spin Ref.[2][3][5] | Experimental Intensity I(%) |
|---|---|---|---|---|---|
| E1 + E2 = 8999.14 keV, Ef = 0 keV ||||||
| 8533.53 | 465.37 | 465.37 | 1/2$^-$ | 1/2$^-$ | 18.427(070) |
| 8121.52 | 878.37 | 878.37 | 1/2$^-$ | 3/2$^-$ | 4.800(072) |
| 7697.51 | 1302.38 | 1302.38 | 1/2$^-$ | 1/2$^-$ | 1.705(095) |
| 6583.49 | 2415.41 | 2415.41 | 3/2⁻ | 3/2⁻ | 0.934(099) |
| 5817.47 | 3181.42 | 3181.42 | 1/2⁻ | 3/2⁺ | 1.290(089) |
| 5435.47 | 3564.43 | 3564.43 | 1/2- | ? | 1.177(096) |
| 5312.46 | 3686.43 | 3686.43 | 1/2⁻ | 3/2⁺ | 0.751(096) |
| 4950.46 | 4049.44 | 4049.44 | 1/2⁻ | ? | 1.147(087) |
| 4284.44 | 4715.45 | 4715.45 | 1/2⁻ | ? | 0.927(099) |
| **E1 + E2 = 8660.04 keV, Ef = 339.10 keV** | | | | | |
| 6105.48 | 2554.41 | 2893.66 | 3/2⁻ | 3/2⁻ | 4.341(070) |
| 5817.47 | 2843.41 | 3181.42 | 3/2⁻ | 3/2⁺ | 5.016(096) |
| 5312.46 | 3347.42 | 3686.43 | 3/2⁻ | 3/2⁺ | 1.280(103) |
| **E1 + E2 = 8533.53 keV, Ef = 465.37 keV** | | | | | |
| 6583.49 | 1950.4 | 2415.41 | 3/2⁻ | 3/2⁻ | 2.474(092) |
| 4858.45 | 3676.43 | 4140.69 | 3/2⁻ | 13/2⁻ | 3.706(076) |
| **E1 + E2 = 3181.42 keV, Ef = 5817.72 keV** | | | | | |
| 2843.41 | 339.1 | 339.1 | 1/2⁻ | 5/2⁻ | 2.367(083) |
| 2717.41 | 465.37 | 465.37 | 1/2⁻ | 1/2⁻ | 2.974(083) |
| 2304.4 | 878.37 | 878.37 | 1/2- | 3/2⁻ | 2.579(086) |
| 1993.4 | 1188.38 | 1188.38 | 1/2⁻ | 5/2⁻ | 5.627(082) |
| 1880.39 | 1302.38 | 1302.38 | 1/2⁻ | 1/2⁻ | 2.237(088) |
| 1735.39 | 1447.39 | 1447.39 | 1/2⁻ | ? | 2.388(107) |
| **E1 + E2 = 2893.66 keV, Ef = 6105.48 keV** | | | | | |
| 2554.41 | 339.1 | 339.1 | 1/2⁻ | 5/2⁻ | 1.739(083) |
| 2016.4 | 878.37 | 878.37 | 1/2⁻ | 3/2⁻ | 3.610(104) |
| 1703.39 | 1188.38 | 1188.28 | 1/2⁻, 5/2⁻ | 5/2⁻ | 0.555(102) |
| **E1 + E2 = 2415.41 keV, Ef = 6583.49 keV** | | | | | |
| 1950.4 | 465.37 | 465.37 | 1/2⁻ | 1/2⁻ | 4.540(108) |
| 1537.39 | 878.37 | 878.37 | 1/2⁻ | 3/2⁻ | 4.611(087) |
| 1226.38 | 1188.38 | 1188.38 | 1/2⁻ | 5/2⁻ | 2.739(106) |

*Note: E1 (keV) is the energy of primary gamma rays, E2 (keV) is the energy of the secondary gamma rays, $E_L$ (keV) is the energy of the intermediate level. Energy fluctuation is about 1 keV. Iγ (%) is intensity of cascade gamma transfer.*

The experimental results measured 26 pairs of cascade gamma transfer with specific energy and intensity. According to the rules of identification of spin and multiple order, spin and parity were calculated and arranged for state that can be measured by experiment, in addition, the spin and parity of uncompleted levels were added in [2].

### 3.2. Gamma transfer probability

From the experimental data of gamma intensity and electromagnetic transfer selection, the magnetic dipole transfer probability was calculated for some levels of $^{59}$Ni nuclear at compound state when capturing neutron. The result also was compared with the data from the single particle model using Equations (3) and (4):

Table 2. The magnetic dipole transfer probability of some levels

| E (keV) | $E_L$ (keV) | $J_i^\pi \to J_f^\pi$ | $T_\gamma^{M1}(Exp)$ | $T_\gamma^{M1}(Th)$ | $\dfrac{T_\gamma^{M1}(Th)}{T_\gamma^{M1}(Exp)}$ |
|---|---|---|---|---|---|
| 2843.41 | 339.10 | 1/2⁻ → 1/2⁻ | 0.576 | 0.580 | 1.006 |
| 2554.41 |  | 1/2⁻ → 1/2⁻ | 0.424 | 0.420 | 0.992 |
| 2717.41 | 465.37 | 1/2⁻ → 1/2⁻ | 0.901 | 0.730 | 0.811 |
| 1950.40 |  | 1/2⁻ → 1/2⁻ | 0.099 | 0.270 | 2.717 |
| 2304.40 | 878.37 | 1/2⁻ → 1/2⁻ | 0.239 | 0.508 | 2.129 |
| 2016.40 |  | 1/2⁻ → 1/2⁻ | 0.334 | 0.341 | 1.019 |
| 1537.39 |  | 1/2⁻ → 1/2⁻ | 0.427 | 0.151 | 0.354 |
| 1993.40 | 1188.38 | 1/2⁻ → 1/2⁻ | 0.631 | 0.539 | 0.854 |
| 1703.39 |  | 1/2⁻ → 1/2⁻ | 0.062 | 0.336 | 5.399 |
| 1226.38 |  | 1/2⁻ → 1/2⁻ | 0.307 | 0.125 | 0.408 |

*Note: E (keV) gamma energy, $E_L$ (keV) energy level, $J_i^\pi$, $J_f^\pi$: spin and parity of the initial and final state, $T_\gamma^{M1}(Exp)$: the experimental magnetic dipole transfer probability. $T_\gamma^{M1}(Th)$: the theoretical magnetic dipole transfer probability.*

Results in Table 2 showed that the ratio between theoretical and experimental transfer probability are approximate to 1, we concludes that $^{59}$Ni nuclei is sufficient with single particle model.

### 4. CONCLUSION

By the empirical study of the cascade transfers of $^{59}$Ni nuclei from $^{58}$Ni(n, 2γ)$^{59}$Ni reaction, we measured 26 pairs of cascade transfer and arranged into scheme, in addition, the relative intensity of the transfers were produced. Using the rules of calculation of spin and parity, the possible spin and parity were calculated for experimental levels, complement for uncompleted levels.

The results also showed the agreement between the calculation from single particle model of nuclei and experiment data. Thus, we conclude that $^{59}$Ni nuclei can be explained by single particle model.